\documentclass[twocolumn,showpacs,preprintnumbers,amsmath,amssymb]{revtex4}
\usepackage{graphicx}
\usepackage{dcolumn}
\usepackage{bm}
\usepackage{color}
\usepackage{subfigure}

\begin{document}
\title{Generation of multi-photon Fock states by bichromatic adiabatic
 passage: \\topological analysis}
\author{M. Amniat-Talab }
\altaffiliation[Also at ]{Physics department, Faculty of Sciences,
Urmia University, P.B. 165, Urmia, Iran.}
\email{amniyatm@u-bourgogne.fr}
\author{S. Lagrange}
\author{S. Gu\'{e}rin}
\email{sguerin@u-bourgogne.fr}
\author{H.R. Jauslin}
\email{jauslin@u-bourgogne.fr}
 \affiliation{Laboratoire
de Physique, UMR CNRS 5027, Universit\'{e} de Bourgogne, B.P.
47870, F-21078 Dijon, France.}
\date{\today}

\begin{abstract}
We propose a robust scheme to generate multi-photon Fock states in
an atom-maser-cavity system using adiabatic passage techniques and
topological properties of the dressed eigenenergy surfaces. The
mechanism is an exchange of photons from the maser field into the
initially empty cavity by bichromatic adiabatic passage. The
number of exchanged photons  depends on the design of the
adiabatic dynamics through and around the conical intersections of
dressed eigenenergy surfaces.
\end{abstract}
 \maketitle

\section{\label{int}Introduction}
Over the past few years, new sources of  anti-bunched light that
are able to emit a single  photon in a given time interval has
been the subject of an intense theoretical and experimental
research. The driving force behind the development of these
non-classical sources is a range of novel applications in quantum
information theory which builds on the laws of quantum mechanics
to transmit, store, and process information in varied and powerful
ways. Advances in this field rely on the ability to manipulate
coherently isolated quantum objects while eliminating incoherent
interactions with the surrounding environment. Single photon
states act as elementary quantum bits (qubits) in quantum
cryptography \cite{jennewein,naik,tittel} and teleportation of a
quantum state \cite{bouwmeester} where their entangled states
enable secure transmission of information.

Many different types of single-photon sources have been proposed
and realized using the controlled excitation of single molecules
\cite{brunelPRL,lounisNature} or of single nitrogen-vacancy
centers in diamond nanocrystals \cite{beveratos2002}, controlled
injection of carriers into a mesoscopic quantum well
\cite{kim1999} and using pulsed excitation of semiconductor
quantum dots \cite{santori,michler}.

In the context of cavity QED, single-photon Fock states have been
produced by a Rabi $\pi$-pulse in a microwave cavity
\cite{maitrePRL,varcoewalther} and by the STIRAP technique in an
optical cavity \cite{hennrichPRL2000} based on the scheme proposed
in \cite{parkins1995} where the Stokes pulse is replaced by a mode
of a high-Q cavity. The STIRAP process has also been studied in a
system of four-level atom interacting with a cavity mode and two
laser pulses, with a coupling scheme which generates two
degenerate dark states \cite{unanyan}. In all these cavity QED
schemes one atom interacts with  a single-mode high-Q cavity and
generates one photon. As the atoms pass through the cavity one by
one, more photons can be added to the cavity. Recently
 another scheme has been proposed to generate a
two-photon Fock state by a single two-level
 atom interacting with  a superconducting cavity which sustains two non-degenerate orthogonally
  polarized modes \cite{bertet2002}.
 The photons are transferred from the source mode into the target mode of the cavity by a
 third-order Raman process. However this scheme is not robust relative to the velocity of
atoms.

 In this paper we propose a scheme in which a two-level
 atom interacts counterintuitively \cite{vitanov} with  a single-mode high-Q cavity
 and a delayed maser field that are both near-resonant with
 the atomic transition, allowing to
 produce a controlled number of photons in the cavity depending
 on the  design of the adiabatic passage. This process is referred to as a bichromatic adiabatic passage since
 two near-resonant interacting fields act on a single transition.
 A related work involving exchange of photons between two laser
 fields through a bichromatic process can be found in Ref.
 \cite{guerin2}.
  The transfer of photons from
 the maser field into the  cavity field is based on the adiabatic passage between two  dressed
 states which are the eigenstates of the coupled atom-maser-cavity system. This process is robust
 because it does not depend on the precise velocity of the atom or on
 the precise tuning of the maser and the cavity frequencies. The dynamics of
the process, under the adiabatic conditions, can be described
completely by the topology of the dressed eigenenergy
 surfaces. This topological aspect is the key to the robustness of
 the process. Our method is based on the calculation of the
 dressed eigenenergy surfaces of the effective Hamiltonian as a
 function of the two Rabi frequencies associated to the maser and the cavity fields, and the application of adiabatic
 principles to determine the dynamics of the process in view of the
 topology of the surfaces.

The paper is structured as follows. In Sec. II, we use the Floquet
formalism and the phase representation of the creation and
annihilation operators to construct the effective Hamiltonian of
the atom-maser-cavity system.  Eigenenergy surfaces of the
effective Hamiltonian are displayed in Sec. III as a function of
the normalized Rabi frequencies of the cavity and the maser
fields. We demonstrate  how the  analysis of these surfaces allows
to design different adapted adiabatic paths leading to different
photon transfers into the cavity field without changing the atomic
population at the end of the interaction. Sec. IV is devoted to
the numerical simulation of the evolution governed by the
effective Hamiltonian and the final probabilities of the one and
two-photon transfer states. Finally, in Sec. V we give some
conclusions and indicate conditions  for experimental
implementation.

\section{\label{cons}Construction of the effective hamiltonian}
We consider a two-level atom of upper and lower states $|+\rangle$
and  $|-\rangle$ and of energy difference $E_{+}-E_{-}=\omega_{0}$
as represented in Fig. 1. We use atomic units in which $\hbar=1$.
The atom in its lower state is released from a source of atoms and
falls through a high-Q cavity with  velocity $v$. The atom first
encounters the vacuum mode of the cavity with frequency
$\omega_{C}$ and waist $W_{C}$ and then the maser beam with
frequency $\omega_{M}$ and waist $W_{M}$. Both the maser and the
cavity fields are near-resonant with the atomic transition. The
distance between the crossing points of the cavity and the maser
axis with the atomic trajectory is $d$. The travelling atom
encounters time dependent and delayed Rabi frequencies of the
cavity and the maser fields:
\begin{eqnarray}\label{rabi}
    G(t)&=&-\mu
    \sqrt{\frac{\omega_{C}}{2\epsilon_{0}V_{\text{mode}}}}~~e^{-(\frac{vt}{W_{C}})^{2}},\nonumber\\
    \Omega(t)&=&-\mu\mathcal{E}_{M}~~e^{-(\frac{vt-d}{W_{M}})^{2}},
\end{eqnarray}
where $\mu$, $V_{\text{mode}}$, $\mathcal{E}_{M}$ are respectively
the dipole moment of  the atomic transition, the effective volume
of the cavity mode and the amplitude of the maser field.
\begin{figure}\label{level}
\centerline{\subfigure{\includegraphics[width=3cm]{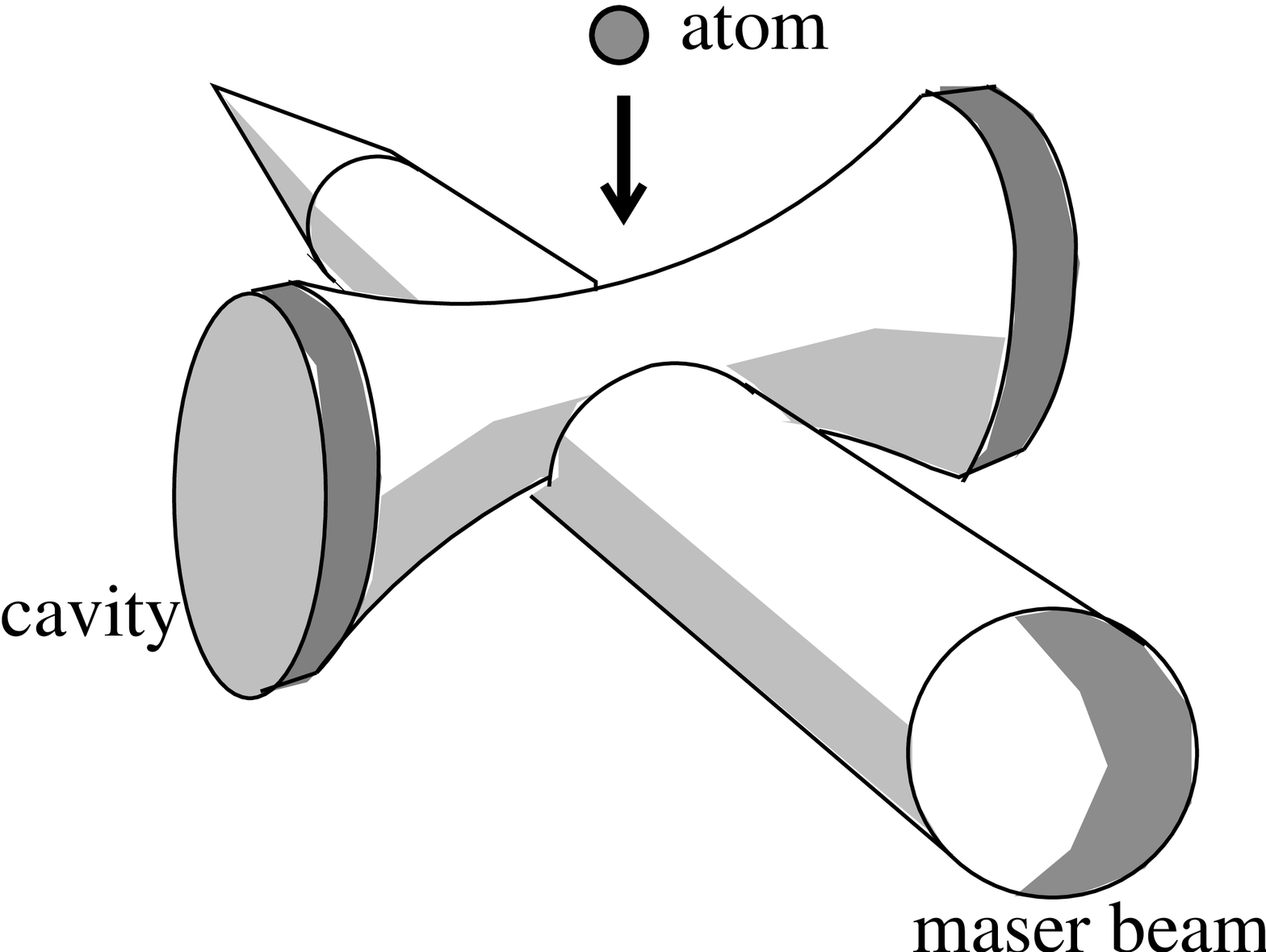}}
\subfigure{
\includegraphics[width=3cm]{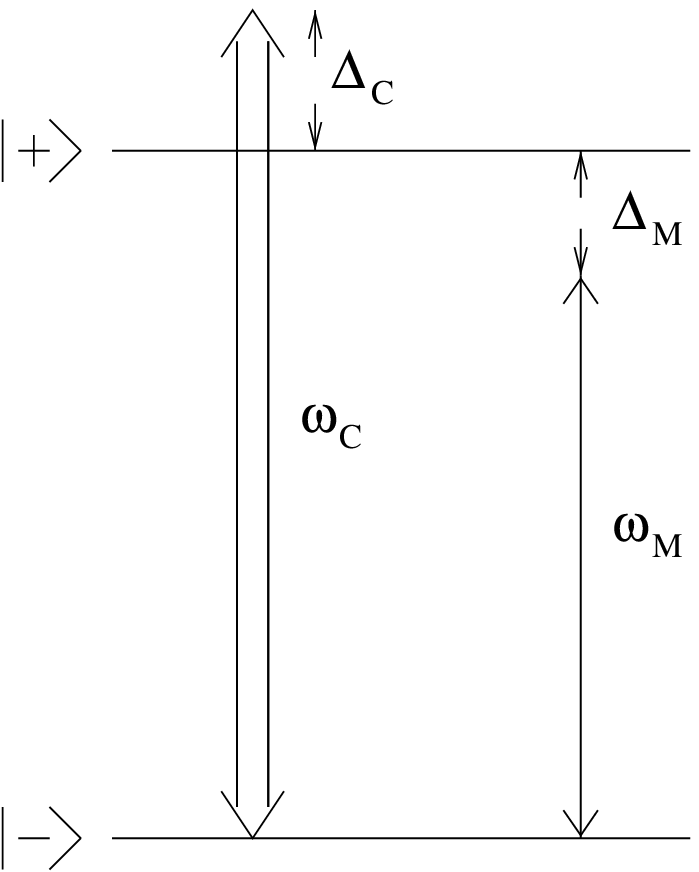}}}
 \caption{ Experimental configuration and  level scheme of the atom.}
\end{figure}
The detuning of the maser and the cavity fields from the atomic
transition are $\Delta_{i}=\omega_{0}-\omega_{i},~~i=C,M$. We take
the frequencies of the fields such that their difference
$\delta=\omega_{C}-\omega_{M}$ is positive and very small with
respect  to $\omega_{C}$ and $\omega_{M}$. Also we assume
\begin{equation}\label{rabicondition}
    \text{max}\{|G(t)|~,~|\Omega(t)|\}\ll \omega_{0}.
\end{equation}
Under these conditions, the counter-rotating terms can be
discarded in the rotating-wave approximation (RWA). The
semiclassical Hamiltonian of the atom-maser-cavity system can thus
be written as
\begin{eqnarray}\label{ham}
H(t) &=&
     \omega_{C}a^{\dag}a
    ~\openone_{2}+\left(%
\begin{array}{cc}
  \omega_{0} & 0 \\
  0 & 0 \\
\end{array}%
\right)
    +G(t)\left(%
\begin{array}{cc}
  0 & a \\
  a^{\dag} & 0 \\
\end{array}%
\right)\nonumber\\
&+&\Omega(t)/2\left(%
\begin{array}{cc}
  0 & e^{-i(\theta_{0}+\omega t)} \\
  e^{i(\theta_{0}+\omega t)} & 0 \\
\end{array}%
\right),
\end{eqnarray}
where $a,a^{\dag}$ are the annihilation and creation operators
 of the cavity field, $\openone_{2}$ is the $2\times 2$ identity matrix and the phase
$\theta_{0}$ is the initial phase of the maser field. The energy
of the lower atomic state has been taken as $0$. This Hamiltonian
acts on the Hilbert space $\mathcal{H} \otimes \mathcal{F}$ where
$\mathcal{H}=\mathbb{C}^{2}$ is the Hilbert space of the atom
generated by $|\pm\rangle$ and $\mathcal{F}$ is the Fock space of
the cavity mode generated by the orthonormal basis
$\{|n\rangle~;~n=0,1,2,\cdots\}$ with $n$ the photon number of the
cavity field. The dynamics of system is determined by the
Schr\"{o}dinger equation
\begin{equation}\label{dyn1}
    i\frac{\partial}{\partial t}\phi(t)=H(t)\phi(t),
\end{equation}
where $\phi(t) \in \mathcal{H} \otimes \mathcal{F}$ with initial
condition $\phi_{0}=\phi(t_{0})=|-\rangle\otimes|n=0\rangle$.
 We can think of Eq.(\ref{dyn1}) as a family of equations parameterized by the phase
$\theta_{0}$. The time-dependent Hamiltonian (\ref{ham}) contains
two different time scales: the period $T=2\pi/\omega$
characterizing fast oscillations  of the maser field and
$T_{int}\approx \frac{W_{M}}{v}\approx \frac{W_{C}}{v}$
characterizing the  slow change of the field amplitudes
$G(t),\Omega(t)$. The fast periodic time-dependence can be taken
into account by use of Floquet theory. The evolution equation in
the Floquet representation  reads:
\begin{equation}\label{dyn2}
i\frac{\partial}{\partial
    t}\psi(t;\theta)=K(t;\theta)\psi(t;\theta),
\end{equation}
where $\theta$ appears as a dynamical variable defined on the unit
circle $\mathbb{S}^{1}$ of length $2\pi$, and $K(t;\theta)$ is the
Floquet Hamiltonian
\begin{eqnarray}\label{Fham}
 K(t;\theta)&=& \omega_{C}a^{\dag}a
    ~\openone_{2}+\left(%
\begin{array}{cc}
  \omega_{0} & 0 \\
  0 & 0 \\
\end{array}%
\right)
-i\omega_{M}\frac{\partial}{\partial\theta}~\openone_{2}\nonumber\\
    &+&G(t)\left(%
\begin{array}{cc}
  0 & a \\
  a^{\dag} & 0 \\
\end{array}%
\right)
+\Omega(t)/2\left(%
\begin{array}{cc}
  0 & e^{-i\theta} \\
  e^{i\theta} & 0 \\
\end{array}%
\right),
\end{eqnarray}
The only time dependence of the Floquet Hamiltonian is from the
slow variation of the field amplitudes. This Hamiltonian acts on
the enlarged Hilbert space \cite{sambe}
 $\mathcal{K}=\mathcal{H}\otimes  \mathcal{F}\otimes\mathcal{L}$
where $\mathcal{L}:=L_{2}(\mathbb{S}^{1}, d\theta/2\pi)$ denotes
the space of square integrable functions on the circle
$\mathbb{S}^{1}$ of length $2\pi$, with a scalar product
\begin{equation}\label{spro}
    \langle
    f_{1}|f_{2}\rangle_{\mathcal{L}}:=\int_{0}^{2\pi}\frac{d\theta}{2\pi}~f_{1}^{*}(\theta)f_{2}(\theta).
\end{equation}
This space is generated by the orthonormal basis
$\{e^{ik\theta}~;~k\in \mathbb{Z}\}$. One can interpret $k$
 as the relative photon number with respect to the (large) average photon
 number $\overline{k}$ of the maser field. The operator
$-i\frac{\partial}{\partial\theta}$ can be interpreted as the
relative photon number operator of the maser field \cite{guerin3}.
The eigenvectors of the zero-field Floquet Hamiltonian are
$|\pm,n,k\rangle=|\pm\rangle\otimes|n\rangle\otimes e^{ik\theta}$
which form an orthonormal basis of the enlarged Hilbert space
$\mathcal{K}$. The relation between the solutions of Eqs.
(\ref{dyn1}) and (\ref{dyn2}) is established as follows
\cite{guerin1}: If $\psi(t;\theta)$ is a solution of (\ref{dyn2})
with initial condition $\psi(t_{0};\theta)=\phi_{0}\otimes
\mathbf{1}_{\mathcal{L}}$, then
$\phi(t;\theta_{0}):=\psi(t;\theta_{0}+\omega t)$ is a solution of
(\ref{dyn1}) with the initial condition $\phi(t_{0})=\phi_{0}$.
$\mathbf{1}_{\mathcal{L}}$ is the basis function $e^{ik\theta}$
with $k=0$.  We remark that if at the end of interaction
$t=t_{f}$, the solution of (\ref{dyn2}) has a form
$\psi(t_{f};\theta)=\phi_{f}\otimes e^{ik\theta}$ then the
probability  for the solution of (\ref{dyn1}) to be found
 in the final states $|\pm,n\rangle$, i.e. $|\langle
\phi(t_{f};\theta_{0})|\pm,n\rangle|^{2}=|\langle\phi_{f}|\pm,n\rangle|^{2}$,
will not depend on the  phase $\theta_{0}+\omega t_{f}$ of the
semiclassical Hamiltonian (\ref{ham}).



The evolution of (\ref{dyn2}) due to slow field amplitudes will be
treated  in the enlarged Hilbert space by adiabatic principles. We
first show that the dynamics of (\ref{Fham}) under the bichromatic
interaction can be described by an effective Hamiltonian. We start
by applying to the Floquet Hamiltonian (\ref{Fham}) the unitary
transformation
\begin{equation}\label{RM}
    R=\left(%
\begin{array}{cc}
  e^{-i\theta} & 0 \\
  0 & 1 \\
\end{array}%
\right),
\end{equation}
 which yields
\begin{eqnarray}\label{KdresL}
    K'&=&R^{\dag}KR=\omega_{C}a^{\dag}a~\openone_{2}-i\omega_{M}\frac{\partial}{\partial\theta}~
    \openone_{2}\nonumber\\
    &+&\left(%
\begin{array}{cc}
  \Delta_{M} & \Omega(t)/2 \\
  \Omega(t)/2 & 0 \\
\end{array}%
\right)+G(t)\left(%
\begin{array}{cc}
  0 & a~e^{i\theta} \\
  a^{\dag}e^{-i\theta}  & 0 \\
\end{array}%
\right).
\end{eqnarray}
The third term of $K'$ , denoted $H_{RWA}$, is the so-called RWA
Hamiltonian, associated to the maser field and the atom. Its
eigenvalues are
$2\lambda_{\pm}^{(0)}=\Delta_{M}\pm\sqrt{(\Delta_{M})^{2}+(\Omega)^{2}}$.
To simplify and decouple the Hamiltonian (\ref{KdresL}), we use
the phase representation of $a$ and $a^{\dag}$ as formulated by
Bialynicki-Birula \cite{birula}:
\begin{equation}\label{phaserep}
    a\rightarrow e^{-i\varphi}~\sqrt{-i\frac{\partial}{\partial\varphi}},~~a^{\dag}\rightarrow\sqrt{-i\frac{\partial}
    {\partial\varphi}}~e^{+i\varphi},~~a^{\dag}a\rightarrow -i\frac{\partial}{\partial\varphi}
\end{equation}
which gives
\begin{eqnarray}\label{Kphase}
K'&=&-i\omega_{C}\frac{\partial}{\partial\varphi}~\openone_{2}-i\omega_{L}\frac{\partial}{\partial\theta}~\openone_{2}+H_{RWA}\nonumber\\
&+&G(t)\left(%
\begin{array}{cc}
  0 & e^{+i(\theta-\varphi)}~\sqrt{-i\frac{\partial}{\partial\varphi}} \\
  \sqrt{-i\frac{\partial}{\partial\varphi}}~e^{-i(\theta-\varphi)} & 0 \\
\end{array}%
\right),
\end{eqnarray}
 Defining the new variables
\begin{equation}\label{newvar}
    \zeta:=\varphi-\theta,~~~~\eta:=\theta,
\end{equation}
 we have
\begin{equation}\label{newpartial}
    \frac{\partial}{\partial\varphi}=\frac{\partial}{\partial\zeta},~~~~\frac{\partial}{\partial\theta}
    =\frac{\partial}{\partial\eta}-\frac{\partial}{\partial\zeta}.
\end{equation}
The eigenbasis of
($-i\frac{\partial}{\partial\varphi}-i\frac{\partial}{\partial\theta}$)
is
$\{e^{in\varphi}e^{ik\theta}~;~n=0,1,2,\cdots~,~k=0,\pm1,\pm2,\cdots\}$
 which can be written as
 \begin{equation}\label{nbasis}
    e^{in\varphi}e^{ik\theta}=e^{in(\zeta+\eta)}e^{ik\eta}=e^{in\zeta}e^{i(n+k)\eta}=e^{in\zeta}e^{im\eta}
\end{equation}
where $m:=n+k=0,\pm1,\pm2,\cdots$. Substituting (\ref{newpartial})
in (\ref{Kphase}) gives
\begin{eqnarray}\label{Knewphase}
K'&=&-i(\omega_{C}-\omega_{M})\frac{\partial}{\partial\zeta}\openone_{2}-i\omega_{M}\frac{\partial}
{\partial\eta}\openone_{2}+H_{RWA}\nonumber\\
&+&G(t)\left(%
\begin{array}{cc}
  0 &  e^{-i\zeta}~\sqrt{-i\frac{\partial}{\partial\zeta}}\\
 \sqrt{-i\frac{\partial}{\partial\zeta}}~e^{+i\zeta} & 0 \\
\end{array}%
\right).
\end{eqnarray}
We can define new operators as
\begin{equation}\label{bop}
    b:=e^{-i\zeta}~\sqrt{-i\frac{\partial}{\partial\zeta}},~~~~b^{\dag}:=\sqrt{-i\frac{\partial}
    {\partial\zeta}}~e^{+i\zeta},
\end{equation}
which verify the standard commutation relations $[b,b^{\dag}]=1$.
The new bosonic operator $b$ that corresponds to the process of
creation of a cavity photon and associated annihilation of a maser
photon, can be intuitively interpreted as the transformation of a
maser photon into a cavity photon. The Hamiltonian
(\ref{Knewphase}) can thus be expressed as
\begin{equation}\label{Kb}
K'=-i\omega_{M}\frac{\partial}{\partial\eta}\openone_{2}+H^{\text{eff}},
\end{equation}
where $H^{\text{eff}}$ is the reduced effective Hamiltonian
\begin{equation}\label{Heff}
H^{\text{eff}}(t)=\delta b^{\dag}b~\openone_{2}+\left(%
\begin{array}{cc}
  \Delta_{M} & \Omega(t)/2 \\
  \Omega(t)/2  & 0 \\
\end{array}%
\right)+G(t)\left(%
\begin{array}{cc}
  0 & b \\
  b^{\dag} & 0 \\
\end{array}%
\right).
\end{equation}
  $K'$ is defined on the Hilbert space generated by the
orthonormal basis $\{|\pm\rangle\otimes e^{in\zeta}\otimes
e^{im\eta}~~;~~n=0,1,2,\cdots ;~ m=0,\pm1,\pm2,\cdots\}$ and
$H^{\text{eff}}$ is defined on the Hilbert space generated by the
orthonormal basis $\{|\pm\rangle\otimes
e^{in\zeta}~~;~~n=0,1,2,\cdots\}$ where $n$ is the number of
exchanged photons from the maser field into the cavity field.

\section{\label{surftop}Topology of the dressed eigenenergy surfaces}
The dressed eigenenergy surfaces of $K'$ (\ref{Kb}) can be
calculated numerically and can be displayed as a function of the
normalized Rabi frequencies $G/\delta$ and $\Omega/\delta$. These
surfaces are grouped in  families for different values of $m$,
each of which  for zero fields consists an infinite set of
eigenvalues with equal spacing $\Delta_{M}$. In what follows, we
study the $m=0$ family only which means $k=-n$ ($k\in \mathbb{Z}$
is the relative photon number of the maser field and $n\geq 0$ is
the photon number of the cavity field). The labelling of the
dressed eigenenergy surfaces can be performed in terms of the
eigenvectors of the zero-field original Hamiltonian,
\begin{equation}\label{zfHam}
    K'(\Omega=0,G=0)=(\delta
    b^{\dag}b-i\omega_{M}\frac{\partial}{\partial\eta})\openone_{2}
    +\left(%
\begin{array}{cc}
   \Delta_{M}& 0 \\
  0 & 0 \\
\end{array}%
\right),
\end{equation}
with eigenvalues
\begin{eqnarray}
 E'_{+,n,m}(\Omega=0,G=0)&=&\delta n+m\omega_{M}+\Delta_{M},\nonumber\\
 E'_{-,n,m}(\Omega=0,G=0)&=&\delta n+m\omega_{M}.
\end{eqnarray}
Since the eigenvectors of $K$ and $K'$ are related by the
transformation (\ref{RM}) as $|\varphi\rangle=R|\varphi\rangle'$,
the correspondence between the eigenvalues of the zero-field
effective Hamiltonian and the eigenvectors of the original
zero-field Hamiltonian are
\begin{eqnarray}
  E'(\Omega=0,G=0)_{+,n,m=0} &\Leftrightarrow& |+,n\rangle'= |+,n,-n-1\rangle \nonumber\\
  E'(\Omega=0,G=0)_{-,n,m=0} &\Leftrightarrow&
  |-,n\rangle'=|-,n,-n\rangle.\nonumber\\
\end{eqnarray}

Figure 2 represents the $m=0$ family of the eigenenergy surfaces
of $K'$  as function of the instantaneous normalized Rabi
frequencies $G/\delta$ and $\Omega/\delta$. Any two neighboring
surfaces have conical intersections on the plane $G=0$ and also on
the plane $\Omega=0$ (except the first surface), corresponding to
the situations where only one of the fields (maser or cavity)  is
interacting with the atom. The topology of these surfaces,
determined by the conical intersections, presents insight into the
various atomic population and photon transfers from the maser
field into the cavity field that can be produced by designing an
appropriate path connecting the initial and the the chosen final
states. Each path corresponds to a choice of the envelope of the
pulses. In the adiabatic limit, when the pulses vary sufficiently
slowly,  the solution of the time-dependent dressed
Schr\"{o}dinger equation follows the instantaneous dressed
eigenvectors,  following the path on the surface that is
continuously connected to the initial state. We start with the
dressed state $|-,0,0\rangle$, i.e., the lower atomic state with
zero photons in the cavity field. Its energy is shown in Fig. 2 as
the starting point of the various paths.  The paths shown in Fig.
2 describe accurately the dynamics if the time dependence of the
envelopes is slow enough according to the Landau-Zener \cite{L,Z}
and Dykhne-Davis-Pechukas \cite{D,DP} analysis. If two (uncoupled)
eigenvalues cross, the adiabatic theorem of Born and Fock
\cite{BF} shows that the dynamics follows diabatically the
crossing. This implies that the various dynamics shown in Fig. 2
are a combination of a global adiabatic passage around the conical
intersections and local diabatic evolutions through (or in the
neighborhood) of conical intersections of the eigenenergy surfaces
\cite{yatsenko}.
\begin{figure}
  \includegraphics[width=7.5cm]{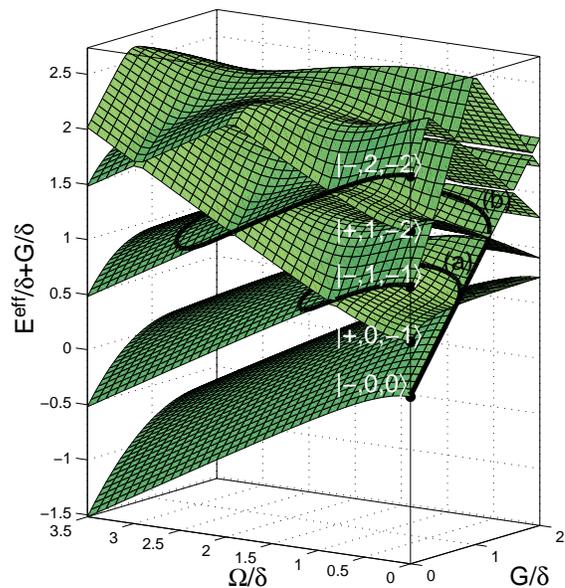}\\
  \caption{First five eigenenergy surfaces (in units of $\delta$) of $H^{\text{eff}}$ as functions of $G$
   and $\Omega$ for $\delta=2\Delta_{M}=-2\Delta_{C}$. The term $G$ has been added to $E^{\text{eff}}$ for clarity of
    display. The solid paths(a,b)
correspond to  adiabatic evolutions which start
   from $|-,0,0\rangle$ state and end at $|-,1,-1\rangle$ and  $|-,2,-2\rangle$ respectively.
   }\label{onefot}
\end{figure}

We consider the action of two smooth pulses, associated with the
Rabi frequencies $G(t)$ and $\Omega(t)$, which act on the
two-level atom with a time delay $\tau=d/v$. Figure 2 shows two
examples of the adiabatic paths of different  peak amplitudes of
the Rabi frequencies and leading to two different photon transfers
into the cavity field without changing the atomic population at
the end of the interaction. Each of the two black paths (labelled
(a) and (b)) corresponds to a sequence of two smooth pulses, shown
in Fig. 3(a) and (c), of equal length $T_{int}$ and different peak
Rabi frequencies $\Omega_{\text{max}},G_{\text{max}}$ , separated
by a delay such that the cavity pulse interacts before the maser
pulse.

For the path (a), the dynamics goes through the first intersection
(on the $\Omega=0$ plane) between the first and the second
surfaces, but not the second intersection between the second and
the third surface. The crossing of the first intersection as $G$
increases with $\Omega=0$, brings the dressed system into the
second eigenenergy surface. Turning on and increasing the
amplitude $\Omega$ (while $G$ decreases) moves the path across
this surface. When the maser field decreases, the curve crosses
another intersection between the second and the third surface
(with $G=0$) that brings the system to the third surface, on which
the path (a) stays until the end of the pulse $\Omega$. The
transfer state is finally connected to the state $|-,1,-1\rangle$:
there is no final transfer of atomic population, but one
$\omega_{M}$ photon has been absorbed from the maser field and one
$\omega_{C}$ photon has been emitted into the cavity field at the
end of the process. The path (b) allows the dynamics (on the
$\Omega=0$ plane) to go through the second intersection, but not
the third intersection. The next two intersections of the path (b)
are located (on the plane $G=0$) between the third and the fourth
and between the fourth and the fifth eigenenergy surfaces and the
system is finally connected to the state $|-,2,-2\rangle$: there
is again no final transfer of atomic population, but two
$\omega_{M}$ photons have been absorbed from the maser field and
two $\omega_{C}$ photons have been emitted into the cavity field
at the end of the process. If the peak amplitudes are taken even
larger such that $n$ conical intersections (dynamical resonances)
are crossed when $G$ rises with $\Omega=0$ and then $n$
intersection are crossed when $\Omega$ decreases with $G=0$, the
final state of the system will be $|-,n,-n\rangle$, i.e., the
emission of $n~~\omega_{C}$ photons into the cavity field and the
absorbtion of $n~~\omega_{M}$ photons from the maser field, with
no final atomic population transfer.
\begin{figure}
\includegraphics[width=9cm]{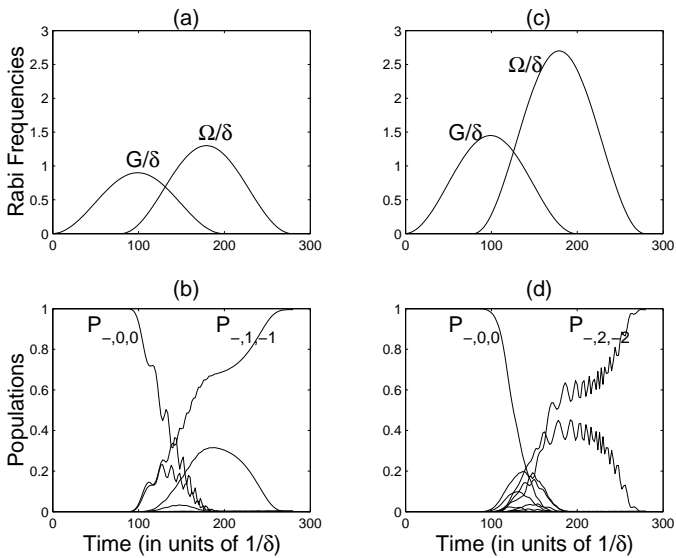}\\
  \caption{(a) and (c): Normalized Rabi frequencies with different values of amplitudes for one-photon and
  two-photon transfers as a function of time. (b) and (d): Time evolution of the populations for one and
  two-photon transfers.}\label{pops}
\end{figure}
The analysis of the eigenenergy surfaces allows one to determine
adapted amplitudes of the Rabi frequencies which will permit to
transfer $n$ photons into the cavity field in a robust way.
\section{\label{numsim}numerical simulation}
The time evolution of the system for each family of the
eigenenergy surfaces is given by the Schr\"{o}dinger equation
\begin{equation}\label{sch}
i\frac{\partial}{\partial t}\Phi(t)=H^{\text{eff}}(t)\Phi(t)
\end{equation}
  The time dependence of the Rabi
frequencies are delayed Gaussians of the form (\ref{rabi}).
Figures 3(a),(c) show the profile of the Rabi frequencies as
functions of time for one-photon and two-photon transfer with an
interaction time (full width at half maximum) $T_{int}=66/\delta$
and a time delay $\tau=57/\delta$. The adapted amplitudes of the
Rabi frequencies for $n$-photon ($n$=1,2) transfer  correspond to
the paths (a) and (b) in Fig. 2. The  condition for global
adiabaticity $|\Delta_{M,C}|T_{int}=\frac{\delta}{2}T_{int}=33\gg
1$ is well satisfied. Figures 3(b,d) present the time evolution of
populations calculated numerically by solving (\ref{sch}). The
atom-maser-cavity system in the initial state $|-,0,0\rangle$ with
the suitable forms of Rabi frequencies (Figs. 3(a,c)) evolves to
the finale states $|-,1,-1\rangle$ and $|-,2,-2\rangle$
respectively with  probabilities of $P_{-,1,-1}=|\langle
-,1,-1|\Phi(t_{f})\rangle|^{2}=0.99$ and $P_{-,2,-2}=|\langle
-,2,-2|\Phi(t_{f})\rangle|^{2}=0.98$.

Fig. 4 displays the contour plot of the final population as a
function of the normalized Rabi frequencies for one and two-photon
transfers. The white regions represent  the adapted values of the
Rabi frequencies for which the final probability of one and
two-photon transfers are maximal. This figure shows that the
bichromatic adiabatic passage is more robust with respect to the
maser Rabi frequency than with respect to the cavity one. The
reason comes back to the special structure of the dressed
eigenenergy surfaces in Fig. 2. We can see that on the $G=0$
plane, between the first surface and the second one there isn't
any intersection and between the $n$th surface and  its
neighboring surfaces there are $(n-1)$  intersections, i.e. after
the ($n-1$)th intersection there are no others. On the other hand
on the $\Omega=0$ plane, as the value of $G$  increases, the
distance between neighboring intersections decreases. In general,
as the distance of  conical intersections between neighboring
surfaces decreases, the robustness of the adiabatic passage is
 also decreased.
\begin{figure}
\includegraphics[width=9cm]{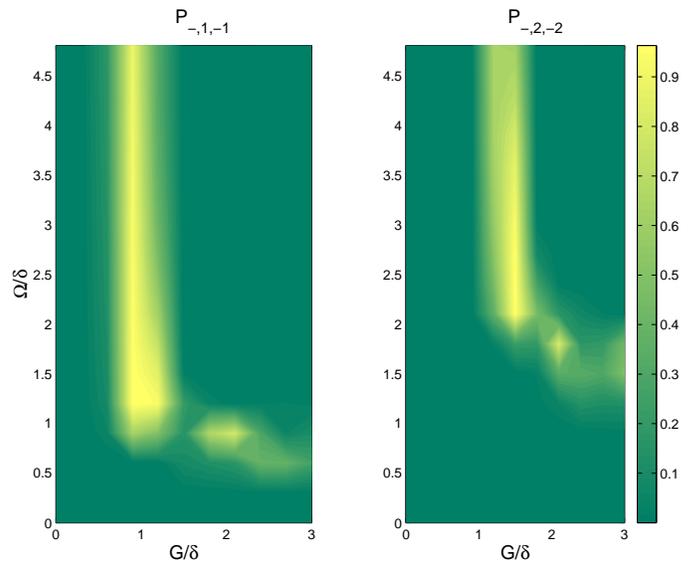}\\
  \caption{ Contour plot of the populations $P_{-,1,-1}$ and $P_{-,2,-2}$ for one-photon and
  two-photon transfers as a function of normalized Rabi
   frequencies.}\label{fourfig}
\end{figure}

\section{\label{con}discussions and conclusions}
Using the topological properties of dressed eigenenergy surfaces
of the effective Hamiltonian of the atom-maser-cavity system, we
have determined adiabatic paths to transfer $n$ photons from the
maser field into the cavity field to generate  a $n$-photon Fock
state. The realization of parameters satisfying the conditions of
the proposed scheme appears feasible with progressive improvements
to experiments with high-Q microwave cavities. In this analysis we
have assumed that the
 interaction time between the two-state atom and the fields is  short compared to the
 cavity lifetime $T_{\text{cav}}$ and the atom's excited state lifetime
 $T_{\text{at}}$, i.e. $T_{\text{int}}\ll T_{\text{cav}} , T_{\text{at}}$, which are essential for an
 experimental setup
 and avoiding  decoherence effects. In the microwave domain, the
 radiative lifetime of circular Rydberg states -- of the order of $T_{\text{at}}=30
 $ms -- are much longer than those for noncircular  Rydberg states. The  typical value of the cavity lifetime is of
 the order of $T_{\text{cav}}=1$ ms (corresponding to $Q=3\times
 10^{8}$) and the  upper limit of interaction time is  $T_{\text{int}}=100$ $\mu$s (atom with a velocity
 of 100 m/s with the cavity mode waist of $W_{C}=6$ mm)
 \cite{raimond}.

 The condition of global adiabaticity $G_{\text{max}}~T_{\text{int}}\gg
 1$ for the typical value of $G_{\text{max}}\approx 0.15$ MHz \cite{raimond} is
 well satisfied $(G_{\text{max}}~T_{\text{int}}\approx 15$).
 Analysis for additional conditions to go diabatically through conical
 intersections can be found in \cite{yatsenko}.
\begin{acknowledgments}
M. A-T gratefully acknowledges the financial support of the French
Society SFERE and the MSRT of Iran. We acknowledge support  from
the Conseil R\'{e}gional de Bourgogne and the Centre de Ressources
Informatiques de l'Universit\'{e} de Bourgogne.
\end{acknowledgments}

\end{document}